\documentstyle[aps,prb,preprint,amsmath,graphics,amssymb,fontenc]{revtex}

\makeatletter

\providecommand{\LyX}{L\kern-.1667em\lower.25em\hbox{Y}\kern-.125emX\@}
\let\SF@@footnote\footnote
\def\footnote{\ifx\protect\@typeset@protect
    \expandafter\SF@@footnote
  \else
    \expandafter\SF@gobble@opt
  \fi
}
\expandafter\def\csname SF@gobble@opt \endcsname{\@ifnextchar[
  \SF@gobble@twobracket
  \@gobble
}
\edef\SF@gobble@opt{\noexpand\protect
  \expandafter\noexpand\csname SF@gobble@opt \endcsname}
\def\SF@gobble@twobracket[#1]#2{}

\makeatother
\draft
\begin{document}

\title{Peak positions and shapes in neutron pair correlation functions from powders
of highly anisotropic crystals}

\author{D. A. Dimitrov, H. R\"oder, and A. R. Bishop}

\address{Los Alamos National Laboratory, Los Alamos, New Mexico 87545}
\date{\today}
\maketitle

\begin{abstract}
The effect of the powder average on the peak shapes and positions in neutron
pair distribution functions of polycrystalline materials is examined. It is
shown that for highly anisotropic crystals, the powder average leads to shifts
in peak positions and to non-Gaussian peak shapes. The peak shifts can be as
large as several percent of the lattice spacing. 
\end{abstract}


\narrowtext

\section{Introduction}

\label{Introduction}

The purpose of this paper is to demonstrate that the peak shapes in the powder
averaged static pair distribution functions (PDF) of highly anisotropic \emph{harmonic}
crystals are generally non-Gaussian and the peak positions (maxima) are at different
distances than the average atomic pair distances. Our approach is to consider
the effect of the powder average on specific peaks of the theoretical PDF. We
discuss the relevance of these results to what is actually measured in neutron
diffraction experiments. 

The lattice dynamics properties of highly anisotropic crystals, in particular
for crystalline materials of layered and chain structure, are known to lead
to significant deviations from the Debye approximation for thermodynamic quantities
rendering it inapplicable in these cases. Lifshitz\cite{Lif52} showed that
there are different low temperature regimes in which the specific heat does
not behave as \( T^{3} \) but rather as \( T^{\alpha } \) with the exponent
\( \alpha  \) having different values depending on the different regimes for
the phonon dispersion relations. The behavior of the thermal expansion tensor
is similar. The anisotropy also leads to very different in magnitude Debye-Waller
factors\cite{Kri82} and to violation of the Cauchy relations for the elastic
constants. The latter is due to the inability to describe the inter-atomic forces
with central pair potentials. A number of different models have been developed
for the inter-atomic potentials to deal with the anisotropy, depending on the
physical properties of the crystal concerned. For example, generalized force
constants models are considered for some simple metals, bond bending forces
are introduced for the strongly anisotropic covalent crystals, anisotropic ionic
polarizabilities for ionic crystals, etc. It is of interest to determine the
effects of highly anisotropic harmonic lattice vibrations together with the
powder average on the static pair distribution function (PDF) and the implications
for the corresponding measurements in neutron powder diffraction experiments. 

The space-time pair correlation function \( G\left( {\mathbf{r}},\, t\right)  \)
introduced by Van Hove\cite{Hov54} in the theory of neutron scattering has
been primarily studied, both experimentally and theoretically, for liquids and
amorphous materials.\cite{Was80} The correlation function in liquids is often
assumed to be Gaussian provided the system is considered as isotropic.\cite{Rah62::1,Kap65}
The interpretation of the experimental data is quite simplified by assuming
Gaussian peaks, and indeed much has been learnt about the structure and the
interactions in such systems from the pair correlation function. 

The determination of the static PDF from experimental data on powder samples
is obtained by first extracting the static structure factor \( S(q) \), where
\( q=\left| \mathbf{q}\right|  \), from the measured (effective) coherent cross
section \( \left( \frac{d\sigma _{coh}}{d\Omega }\right) _{eff} \) and then
Fourier transforming it. \( S(q) \) has structure for large momentum transfers
\( \hbar q \) in crystals due to their long-range periodic structure. For an
accurate PDF to be obtained from the Fourier transform of \( S(q) \) in such
materials, the structure factor should be known to large enough values of \( q \).
Only recently the development of high-intensity neutron sources has allowed
the measurement of effective cross sections up to \( q\approx 50 \) \AA\( ^{-1} \)
or larger and thus the ability to obtain a PDF for polycrystalline materials
from neutron diffraction experiments.\cite{Ega90,Tob91} 

The extraction of the structure factor \( S(q) \) from the neutron scattering
data is based on several assumptions. The first Born approximation is assumed
to be applicable for the interpretation of the differential cross section. This
is the case, at least in the static approximation when the energy transfer of
the scattered particle or photon, in the case of x-rays, is negligible compared
to its own energy.\cite{Hov54} 
The static approximation is applicable to the
scattering of x-rays but cannot be applied 
directly to the scattering of neutrons
because of their specific energy and wavelength scales
(note that the static structure factor $S\left( \mathbf{q}\right)$  which is
obtained by integrating out the $\omega$ dependence of 
$S\left( \mathbf{q},\, \omega \right)$ is different from the 
structure factor obtained in the {\em static approximation}\cite{Squ96}).
Corrections to the static
approximation have been developed by Placzek\cite{Pla52} and by Wick\cite{Wic54}
which are applicable for high incident neutron energies. 
Yarnell \emph{et al.}\cite{Yar73}
have shown that the application of the Placzek correction to neutron detectors
with different efficiency can lead to determination of the structure factor
\( S(q) \) with an accuracy of \( \sim 0.01 \) for liquid Ar. However, it
has been pointed out by Ascarelli and Cagliati\cite{Asc66} that there are cases
when the Placzek correction could lead to large errors, particularly for values
of \( q \) for which the curvature of \( S(q) \) is high. Generally, this
requires a careful study of the Placzek correction before its application is
adopted. We shall assume that the Placzek correction is applicable in order
to determine the \( S(q) \) from the differential scattering cross section,
as is usually done when analyzing the experimental data\cite{Ege87} for liquids
as well as for crystals.\cite{Tob91} We also assume that all the other corrections
and normalizations applied to the raw data, such as multiple scattering, incoherent
scattering calibration, absorption, polarization, etc., and measurements to
large momentum transfers are made so that the Fourier transformation termination
errors are very small. A careful experimental determination of the PDF which
satisfies these assumptions will allow for a meaningful comparison with a theoretical
calculation.\cite{Dim99}

In Sec. \ref{StaticPDF}, we introduce the static pair distribution function
for harmonic crystals with arbitrary structure and its relation to the experimental
quantities measured. The powder average for specific cases is considered in
Sec. \ref{PowderAverage} which contains the main results of this paper. 
The implications of these results together with their relevance to 
x-ray experiments are discussed in Sec. \ref{Implications}.

\section{Experimental and Theoretical Background}

\label{StaticPDF}

The coherent differential cross-section for scattering of thermal neutrons in
the first Born approximation is given by\cite{Squ96}

\begin{equation}
\label{DiffCrossSection}
\left( \frac{d^{2}\sigma }{d\Omega dE'}\right) _{coh}=\frac{\sigma _{coh}}{4\pi }\frac{k'}{k}\frac{N}{2\pi \hbar }S\left( \mathbf{q},\, \omega \right) ,
\end{equation}
\begin{equation}
\label{StrFactor}
S\left( \mathbf{q},\, \omega \right) =\frac{1}{N}\sum _{j,j'=1}^{N}\int _{-\infty }^{\infty }\left\langle e^{-i{\mathbf{q}}\cdot {\mathbf{r}}_{j}\left( 0\right) }e^{i{\mathbf{q}}\cdot {\mathbf{r}}_{j'}\left( t\right) }\right\rangle e^{-i\omega t}dt,
\end{equation}
 where \( \sigma _{coh}=4\pi \bar{b}^{2} \) (\( \bar{b} \) is the compositionally
averaged scattering length\cite{Lov85v1}), \( \mathbf{k} \) and \( \mathbf{k}'=\mathbf{k}-\mathbf{q} \)
are the initial and final wave vectors of the scattered neutron, \( {\mathbf{r}}_{j}\left( t\right)  \)'s
are the operators for the position vectors of the \( N \) particles in the
target system in the Heisenberg representation, \( S\left( \mathbf{q},\, \omega \right)  \)
is the dynamical structure factor, and \( \left\langle \ldots \right\rangle  \)
denotes thermal average. We have here assumed that the scattering system consist
of only one type of atoms. In the general case, the different types of atoms
will have different scattering lengths and this should modify Eqs. (\ref{DiffCrossSection})
and (\ref{StrFactor}) accordingly. This case will be considered explicitly
for the static PDF in crystals. 

The Van Hove\cite{Hov54} pair distribution function is related to \( S\left( \mathbf{q},\, \omega \right)  \)
via a space-time Fourier transformation\begin{equation}
\label{VanHoveGrt}
G\left( {\mathbf{r}},\, t\right) =\frac{1}{\left( 2\pi \right) ^{3}}\int S\left( {\mathbf{q}},\, \omega \right) e^{-i\left( {\mathbf{q}}\cdot {\mathbf{r}}-\omega t\right) }d{\mathbf{q}}dt.
\end{equation}
 This function is complex since in the general case the position vector operators
for the particles of the scattering system do not commute at different times.
At equal times, \( t=0 \), these operators commute and \( G\left( {\mathbf{r}},\, 0\right)  \)
is real.

Provided \( S\left( \mathbf{q},\, \omega \right)  \) can be extracted from
the measured differential cross-section, Eq.(\ref{DiffCrossSection}), then
Eq. (\ref{VanHoveGrt}) can be used to obtain \( G\left( {\mathbf{r}},\, t\right)  \).
The pair distribution function has a simple physical meaning and is easier to
understand than \( S\left( \mathbf{q},\, \omega \right)  \). It is not currently
feasible to determine experimentally the dynamical structure factor in such
a domain of the four dimensional \( \left( \mathbf{q},\, \omega \right)  \),
space for the Fourier transformation in Eq. (\ref{VanHoveGrt}) to be applied.
However, in some cases the \( \omega  \) integration can be effectively done
by the neutron detectors. This allows for the determination of the static structure
factor\begin{equation}
\label{Sq::Static}
S({\mathbf{q}})=\int _{-\infty }^{\infty }S\left( {\mathbf{q}},\, \omega \right) d\omega =\int G\left( {\mathbf{r}},\, 0\right) e^{i{\mathbf{q}}\cdot {\mathbf{r}}}d{\mathbf{r}},
\end{equation}
 provided \( \mathbf{q} \) is kept constant in the \( \omega  \) integration.
This is the case, at least, in the static approximation. The neutron detectors
carry out this integration at constant scattering angle \( \theta  \) and the
effective coherent cross-section per unit solid angle is\begin{equation}
\label{Sigma::Eff}
\left( \frac{d\sigma }{d\Omega }\right) ^{eff}_{coh}=\frac{\sigma _{coh}}{4\pi }\frac{N}{2\pi \hbar }\int _{-\infty }^{\omega _{max}}\epsilon (k')\frac{k'}{k}S\left( \mathbf{q},\, \omega \right) d\omega ,
\end{equation}
 where \( \hbar w_{max} \) is the energy of the incoming neutron, \( \epsilon (k') \)
is the detector's energy dependent efficiency, and \( \mathbf{q} \) is a function
of the scattering angle \( \theta  \) via \( q^{2}=k^{2}+{k'}^{2}-2kk'\cos (2\theta ) \).
The problem is to obtain \( S(\mathbf{q}) \), as defined by Eq. (\ref{Sq::Static}),
from the measured effective cross section \( \left( \frac{d\sigma }{d\Omega }\right) ^{eff}_{coh} \)
when the static approximation is not valid. The usual way to tackle this problem
is to apply the Placzek correction as mentioned in Sec. \ref{Introduction},
particularly when the incident neutron energy is high compared to the energy
transfers in the scattering processes. 

For powder samples, the correlation functions do not depend on the angular coordinates
of \( \mathbf{q} \) and \( \mathbf{r} \) since they are effectively averaged
over these variables. Then the relations between \( S(q) \) and \begin{equation}
\label{Rho::r}
\rho (r)=G(r,\, 0)-\delta (r)
\end{equation}
 are \begin{equation}
\label{Sq::via::Rhor}
q\left( S\left( q\right) -1\right) =4\pi \int _{0}^{\infty }\left( \rho \left( r\right) -\rho _{0}\right) r\sin \left( qr\right) dr,
\end{equation}
 \begin{equation}
\label{Rhor::via::Sq}
r\left( \rho \left( r\right) -\rho _{0}\right) =\frac{1}{2\pi ^{2}}\int _{0}^{\infty }\left( S\left( q\right) -1\right) q\sin \left( qr\right) dq,
\end{equation}
 where \( \rho _{0}=N/V \), \( V \) is the volume of the system, and the forward
scattering has been subtracted from \( S(q) \), which is the conventional practice
since it is negligible for scattering angles larger than \( \sim 10^{-2} \)
seconds of arc.\cite{Yar73} The experimental PDF, \( \rho _{exp}(r) \), is
obtained from Eq. (\ref{Rhor::via::Sq}) with the upper limit of the integration
\( q=q_{max} \) determined from the maximum value of the scattering vector
accessible in the measurement. If the assumptions made in Sec. \ref{Introduction}
regarding \( \rho _{exp}(r) \) are valid, we can compare this with a theoretically
derived \( \rho (r) \) which we consider next for harmonic crystals. 

The static pair distribution function, \( \rho (\mathbf{r}) \), in a form suitable
for neutron diffraction from a crystal can be defined by \begin{equation}
\label{PDF:def}
\rho ({\mathbf{r}})=\frac{1}{N_{b}}\sum _{{\mathbf{R}}(l'),d,d'}\! \! \! \! \! \! '\frac{b(0d)b(l'd')}{\bar{b}^{2}}\left\langle \delta \left( {\mathbf{r}}-{\mathbf{r}}\left( 0d;l'd'\right) \right) \right\rangle ,
\end{equation}
 where \( N_{b} \) is the number of basis atoms in the unit cell, \( {{\mathbf{R}}}(l) \)
is the position vector of the origin of the \( l \)th unit cell, \( d,\, d' \)
are basis atom indices, \( b(ld) \) is the scattering length of the atom with
equilibrium position at \( {{\mathbf{R}}}(ld)={{\mathbf{R}}}(l)+{{\mathbf{R}}}(d) \). \( {{\mathbf{R}}}(d) \)
is the position vector of the \( d \)th atom in the basis relative to the origin
of the cell. The \( {\mathbf{r}}(0d;l'd')={\mathbf{r}}(0d)-{\mathbf{r}}(l'd') \)
is the difference between the instantaneous positions of the two atoms. The
instantaneous position of an atom is \( {\mathbf{r}}(ld)={{\mathbf{R}}}(ld)+\mathbf{u}(ld) \),
where \( {\mathbf{u}}(ld) \) is the deviation vector from the equilibrium position
(the time arguments of the two \( \mathbf{u} \) vectors in Eq.(\ref{PDF:def})
are not explicitly written since they are at equal times). The sum excludes the
terms with \( l'=0, \) and \( d=d'. \) For harmonic crystals \begin{eqnarray}
\rho ({\mathbf{r}}) & = & \frac{1}{N_{b}}\sum _{{{\mathbf{R}}}(l'),d,d'}\! \! \! \! \! '\frac{b(0d)b(l'd')}{\bar{b}^{2}}\nonumber \\
 &  & \times \frac{e^{-\frac{1}{2}\left( {\mathbf{r}}-{{\mathbf{R}}}\left( 0d;l'd'\right) \right) \cdot {\mathcal{T}}^{-1}\left( 0d;l'd'\right) \cdot \left( {\mathbf{r}}-{{\mathbf{R}}}\left( 0d;l'd'\right) \right) }}{\sqrt{\left( 2\pi \right) ^{3}\det \left[ {\mathcal{T}}\left( 0d;l'd'\right) \right] }},\label{PDF:HA} 
\end{eqnarray}
 where \( {{\mathbf{R}}}(0d;l'd')={{\mathbf{R}}}(0d)-{{\mathbf{R}}}(l'd') \), and the tensor
\( {\mathcal{T}}(0d;l'd') \) is \begin{equation}
\label{To:Matrices}
{\mathcal{T}}_{\alpha \beta }(0d;l'd')=\left\langle \left( {\mathbf{u}}\left( 0d\right) -{\mathbf{u}}\left( l'd'\right) \right) _{\alpha }\left( {\mathbf{u}}\left( 0d\right) -{\mathbf{u}}\left( l'd'\right) \right) _{\beta }\right\rangle .
\end{equation}
 The result for \( G\left( {\mathbf{r}},\, t\right)  \), \( t\neq 0 \), in such
crystals is given by Van Hove.\cite{Hov54} The thermodynamic averages of the
type \( \left\langle {\mathbf{u}}_{\alpha }\left( ld\right) {\mathbf{u}}_{\beta }\left( l'd'\right) \right\rangle  \)
are easily evaluated once the phonon frequencies \( \omega _{s}({\mathbf{k}}) \)
and polarization vectors \( \boldsymbol {\sigma }_{d}^{s}({\mathbf{k}}) \), where
\( s \) is the branch index, are known for reciprocal space \( {\mathbf{k}} \)-vectors
in the first Brillouin zone\begin{eqnarray}
\left\langle {\mathbf{u}}_{\alpha }\left( ld\right) {\mathbf{u}}_{\beta }\left( l'd'\right) \right\rangle  & = & \frac{\hbar }{2N_{k}\sqrt{M_{d}M_{d'}}}\sum _{{\mathbf{k}},s}e^{i{\mathbf{k}}\cdot \left( {{\mathbf{R}}}\left( l\right) -{{\mathbf{R}}}\left( l'\right) \right) }\nonumber \\
 &  & \boldsymbol {\sigma }_{d\alpha }^{s}({\mathbf{k}})\boldsymbol {\sigma }_{d'\beta }^{s*}({\mathbf{k}})\frac{\left( 2n_{s}\left( {\mathbf{k}}\right) +1\right) }{\omega _{s}\left( {\mathbf{k}}\right) }.\label{uu::Tensor} 
\end{eqnarray}
 Here \( n_{s}\left( {\mathbf{k}}\right) =\left( \exp \left( \beta \hbar \omega _{s}\left( {\mathbf{k}}\right) \right) -1\right) ^{-1} \),
\( N_{k} \) is the number of unit cells in a crystal with periodic boundary
conditions (or \( k \) points in the first Brillouin zone), and \( M_{d} \)
is the mass of the \( d \)th type of atom. Note that \( \left\langle {\mathbf{u}}_{\alpha }\left( ld\right) {\mathbf{u}}_{\beta }\left( ld\right) \right\rangle  \)
is related to the Debye-Waller\cite{Lov85v1} factor via \( 2W_{d}({\mathbf{q}})={\mathbf{q}}\cdot \left\langle {\mathbf{u}}\left( ld\right) {\mathbf{u}}\left( ld\right) \right\rangle \cdot {\mathbf{q}} \),
where \( {\mathbf{q}} \) is a reciprocal space vector. 

As expected, the static PDF for harmonic crystals as a function of the vector
argument \( {\mathbf{r}} \), Eq. (\ref{PDF:HA}), consists of a sum of Gaussian
distributions for each pair of atoms, apart from the scattering length factors.
Each distribution is centered at the pair's average distance \( {{\mathbf{R}}}(0d;l'd') \)
and its second moments with respect to \( {{\mathbf{R}}}(0d;l'd') \) are given
by the elements of the tensor \( {\mathcal{T}}(0d;l'd') \). These moments determine
the width of the peaks in the PDF along any direction in \( {\mathbf{r}} \)-space.
It is clear from Eq. (\ref{uu::Tensor}) that the moments depend on \( {{\mathbf{R}}}\left( l\right) -{{\mathbf{R}}}\left( l'\right)  \)
and the type of atoms in the pair. Since \( \left\langle {\mathbf{u}}_{\alpha }\left( ld\right) {\mathbf{u}}_{\beta }\left( l'd'\right) \right\rangle \rightarrow 0 \)
when \( \left| {{\mathbf{R}}}\left( l\right) -{{\mathbf{R}}}\left( l'\right) \right| \rightarrow \infty  \),
the moments of the distribution will approach \( \left\langle {\mathbf{u}}_{\alpha }\left( 0d\right) {\mathbf{u}}_{\beta }\left( 0d\right) \right\rangle +\left\langle {\mathbf{u}}_{\alpha }\left( 0d'\right) {\mathbf{u}}_{\beta }\left( 0d'\right) \right\rangle  \)
and will become distance independent in this limit.

\section{Powder Average}

\label{PowderAverage}

When the scattering is from a crystalline powder sample, the structure factor
depends only on the magnitude of the scattering vector and is averaged over
its angular coordinates. Then the pair distribution function defined by Eq.
(\ref{VanHoveGrt}), or by the inverse Fourier transform of Eq. (\ref{Sq::Static})
for the static PDF, will depend only on \( r \) and not on the direction of
\( {\mathbf{r}} \). In the latter case, this is equivalent to an angular average
in real space

\begin{equation}
\label{PDF:Powder}
\rho (r)=\int _{S_{\Omega _{r}}}\frac{d\Omega _{r}}{4\pi }\rho ({\mathbf{r}}),
\end{equation}
where \( r\equiv |{\mathbf{r}}| \) and \( S_{\Omega _{r}} \) is the surface
of a sphere with a center at the origin of the coordinate system and a radius
\( r \). The static, angularly averaged, \( S(q) \) and \( \rho (r) \) are
related by Eqs. (\ref{Sq::via::Rhor}) and (\ref{Rhor::via::Sq}). 

Below, we will apply Eq. (\ref{PDF:Powder}) to the different peaks in \( \rho ({\mathbf{r}}) \),
as given for crystals by Eq. (\ref{PDF:HA}), in order to examine the effects
of the powder average on the static PDF. Note that if \( \rho (r) \) is to
be obtained from \( S(q) \) via Eq. (\ref{Rhor::via::Sq}), the integral over
\( q \) must be carried out to infinity rather than to a maximum value \( q_{max} \)
determined by the scattering instrument. In what follows, it is undedstood that
the assumptions made in Sec. \ref{Introduction} hold such that the theoretical
PDF obtained from Eqs. (\ref{PDF:HA}) and (\ref{PDF:Powder}) can be compared
with the experimentally determined. 

We consider first the case of an isotropic tensor \( {\mathcal{T}}(0d;l'd') \)
because it is an important limiting case against which we can compare the case
of a general form of \( {\mathcal{T}} \).

\subsection{Gaussian Behavior of \protect\( r\rho (r)\protect \) Peaks}

The simplest case for which the powder average integral, Eq.(\ref{PDF:Powder}),
can be taken analytically is for an atomic pair with an isotropic \( {\mathcal{T}}(0d;l'd')=t^{2}\mathcal{I} \)
tensor (note that generally in taking the integral in Eq.(\ref{PDF:Powder})
for each pair, the \textbf{\( {\mathbf{r}} \)}-space coordinate system can always
be rotated such that \( {\mathcal{T}}(0d;l'd')={\mathrm{diag}}\left\{ t^{2}_{1},\, t^{2}_{2},\, t^{2}_{3}\right\}  \)),
where \( \mathcal{I} \) is the identity tensor. The contribution \emph{\( \rho _{(0d;l'd')}(r) \)}
of such a pair to the total PDF is\begin{eqnarray}
\rho _{(0d;l'd')}(r) & = & \frac{b(0d)b(l'd')}{\bar{b}^{2}4\pi rR(0d;l'd')}(p_{N\left( R(0d;l'd'),\, t^{2}\right) }\left( r\right) \nonumber \\
 &  & -p_{N\left( -R(0d;l'd'),\, t^{2}\right) }\left( r\right) ),\label{Isotr:Peak} 
\end{eqnarray}
 where \( p_{N\left( R,\, t^{2}\right) }\left( r\right) =e^{-\left( r-R\right) ^{2}/2t^{2}}/\sqrt{2\pi t^{2}} \)
is the density of the normal distribution with a mean equal to \( R \) and
standard deviation \( t \). The contribution of \( p_{N\left( -R(0d;l'd'),\, t^{2}\right) }(r) \)
to the PDF can often be neglected when \( t^{2} \) is large enough, and
for \( r \) of the order of the pair distances in a crystalline solid which
are usually equal to several Angstroms or larger. Then the peak in \( r\rho (r) \)
due to such a pair of atoms will be approximately a Gaussian (centered at the
equilibrium distance \( R(0d;l'd') \) between the atoms and with a standard
deviation equal to \( t \)) \emph{divided} by \( R(0d;l'd') \) (apart from
the constant scattering length factor \( b(0d)b(l'd')/4\pi \bar{b}^{2} \)).
Therefore, if for all pairs \( {\mathcal{T}}\left( 0d;l'd'\right) =t^{2}\left( 0d;l'd'\right) \mathcal{I} \)
were isotropic (which is \emph{not} the case for crystalline solids) the powder
averaged PDF would be\begin{equation}
\label{Isotr:PDF}
\rho (r)\approx \frac{1}{N_{b}}\sum _{{{\mathbf{R}}}(l'),d,d'}\! \! \! \! \! '\frac{b(0d)b(l'd')}{\bar{b}^{2}4\pi r}\frac{p_{N\left( R(0d;l'd'),\, t^{2}(0d;l'd')\right) }(r)}{R(0d;l'd')}.
\end{equation}
 Strictly, \( 4\pi r\rho (r) \) is not a sum of Gaussians (again neglecting
the scattering length factors) since each \( p_{N\left( R(0d;l'd'),\, t^{2}(0d;l'd')\right) }(r) \)
is also scaled by its mean value \( R(0d;l'd') \) which is different for the
pairs nonequivalent by symmetry. The form for an isotropic PDF peak given by
Eq.(\ref{Isotr:Peak}) can be used to approximate peaks in diagonally cubic
crystals\cite{BH54} when \[
\left| \left\langle {\mathbf{u}}_{\alpha }(0d){\mathbf{u}}_{\beta }(l'd')\right\rangle +\left\langle {\mathbf{u}}_{\alpha }(l'd'){\mathbf{u}}_{\beta }(0d)\right\rangle \right| \ll u^{2}(0d)+u^{2}(l'd'),\]
for \( \alpha ,\, \beta =x,\, y,\, z, \) since in such crystals \( \left\langle {\mathbf{u}}(ld){\mathbf{u}}(ld)\right\rangle =u^{2}(ld)\mathcal{I} \).
This may be applicable to pairs of atoms separated by a large distance relative
to the nearest neighbor distance such that the correlation terms become much
smaller than the diagonal elements of the self correlation ones. This is a special
case of the more general statement that the long-range part of the PDF can be
obtained to a very good approximation only from the Fourier transformation of
the elastic part of \( S(q) \).

\subsection{Non-Gaussian Behavior of \protect\( r\rho (r)\protect \) Peaks}

\begin{figure}

\caption{\label{peak::shapes}Non Gaussian peaks in the normalized distribution of \protect\( r\rho _{R}(r)\protect \)
vs \protect\( r/R\protect \) for two values of \protect\( \left( R,\, t^{-2}_{\bot },\, t^{-2}_{\Vert }\right) \protect \).
The magnitures of \protect\( R\protect \) are given in the inset of the figure.
Each peak position is shifted at a value greater than the corresponding average
distance \protect\( R\protect \). }

{\par\centering \resizebox*{3in}{3in}{\includegraphics{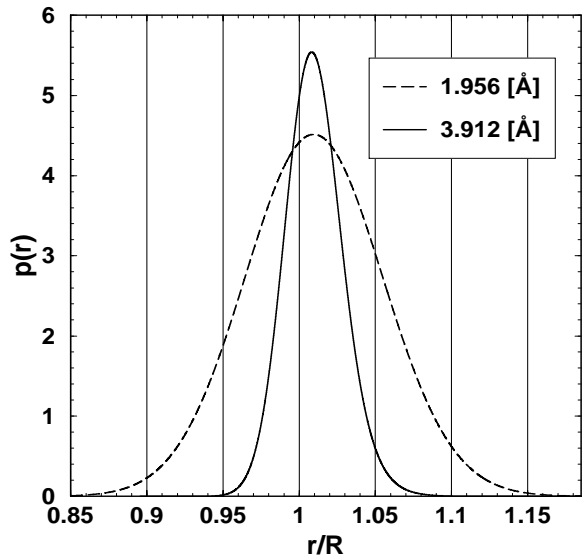}} \par}
\end{figure}

We consider next a case which allows us to study the effect of the powder average
for a strongly anisotropic tensor \( {\mathcal{T}} \) and which can still be
handled analytically. The powder average contribution to the PDF due to a pair
of atoms with \( {\mathcal{T}}(0d;l'd')={\mathrm{diag}}\left\{ t^{2}_{\bot },\, t_{\bot }^{2},\, t_{\Vert }^{2}\right\}  \),
\( t_{\Vert }^{2}<t_{\bot }^{2}, \) and \( {{\mathbf{R}}}(0d;l'd')=(0,\, 0,\, R) \)
is \begin{equation}
\label{AnisotrCase:Peak}
\rho _{(0d;l'd')}\left( r\right) =\frac{1}{4\pi r}\left( p_{N\left( R,t^{2}_{\Vert }\right) }(r)G\left( ar-bR\right) +p_{N\left( -R,t^{2}_{\Vert }\right) }(r)G\left( ar+bR\right) \right) ,
\end{equation}
where\[
G(x)=\sqrt{\pi t^{2}_{\Vert }/2t^{2}_{\bot }\left( t^{2}_{\bot }-t^{2}_{\Vert }\right) }\exp \left( x^{2}\right) \mathrm{erf}\left( x\right) ,\]
 \( a=\sqrt{\left( t_{\bot }^{2}-t_{\Vert }^{2}\right) /2t^{2}_{\bot }t^{2}_{\Vert }} \),
\( b=\sqrt{t^{2}_{\bot }/2t^{2}_{\Vert }\left( t_{\bot }^{2}-t_{\Vert }^{2}\right) } \),
\( \mathrm{erf}(x) \) is the error function, and we have omitted the scattering
lengths scaling factor \( b(0d)b(l'd')/\bar{b}^{2} \). It is straightforward
to check that in the limit \( t_{\Vert }^{2}\rightarrow t_{\bot }^{2} \) from below
we recover the result for the isotropic case given by Eq.(\ref{Isotr:Peak}).
These specific forms of \( {\mathcal{T}} \) and \( {{\mathbf{R}}} \) arise, for
example, for specific \( O_{\mathrm{I}} \)-\( B \) or \( O_{\mathrm{I}} \)-\( O_{\mathrm{I}} \)
nearest neighbor pairs of atoms in \( ABO_{3} \) cubic perovskites. We use
here Cowley's notation\cite{Cow64} for this structure. The self correlation
term \( \left\langle {\mathbf{u}}(B){\mathbf{u}}(B)\right\rangle  \) for the \( B \)
atoms is isotropic and the anisotropy in the \( {\mathcal{T}} \) is due to the
oxygen self correlation term. The latter is proportional to its Debye-Waller
factor and has the same form as \( {\mathcal{T}} \). The \( B \)-\( O \) correlation
terms also contribute to increasing the anisotropy of \( {\mathcal{T}} \). For
the \( O_{\mathrm{I}} \)-\( O_{\mathrm{I}} \) pair of atoms, the \( {\mathcal{T}} \)
tensor is even more anisotropic because of the contributions of both atom's
anisotropic \( \left\langle {\mathbf{u}}(O_{\mathrm{I}}){\mathbf{u}}(O_{\mathrm{I}})\right\rangle  \)
terms. Physically, this is associated with the local environment of the \( O_{\mathrm{I}} \)
atom. It has four \( A \) nearest neighbors (nn) in the \( xy \) plane and
two nn B atoms along the \( z \) axis. The oxygen polarizability is also very
anisotropic and it depends strongly on the oxygen's environment.\cite{Mig76}
To study the peak positions and shapes as functions of \( t_{\bot }^{2} \)
and \( t_{\Vert }^{2} \) we have chosen \( R=1.956 \) and \( R=3.912 \) \AA (the
lattice constant in cubic \( ABO_{3} \) perovskites is often close to 3.9 \AA)
which corresponds to the first nearest neighbor \( O_{\mathrm{I}} \)-\( B \)
and \( O_{\mathrm{I}} \)-\( O_{\mathrm{I}} \) bond distances. The \( t_{\bot }^{2} \),
\( t_{\Vert }^{2} \) are varied in intervals such that \( 5\leq t_{\bot }^{-2}\leq 99 \)
\AA\( ^{-2} \) and \( 100\leq t_{\Vert }^{-2}\leq 250 \) \AA\( ^{-2} \).
Lattice dynamics calculations\cite{DRL::Unpublished} on La\( _{0.7} \)Sr\( _{0.3} \)MnO\( _{3} \)
using the cubic perovskite structure and the nonlinear shell model of Migoni
\emph{et al.\cite{Mig76}} show that these intervals for \( t_{\bot }^{2} \),
\( t_{\Vert }^{2} \) could be physically relevant. The peaks of the normalized
(to unity) distributions \( p(r)=Const\times r\rho _{R}(r) \) for two \( \left( R,\, t^{-2}_{\bot },\, t^{-2}_{\Vert }\right)  \)
sets of parameters, \( \left( 1.956,\, 21.71,\, 133.99\right)  \) and \( \left( 3.912,\, 6.23,\, 248.11\right)  \),
are shown in Fig. \ref{peak::shapes}. These values of of \( \left( R,\, t^{-2}_{\bot },\, t^{-2}_{\Vert }\right)  \)
arise for a specific set of the shell model parameters chosen from a Monte Carlo
sample sequence in a domain of their phase space. The sequence was generated
in a Reverse Monte Carlo estimation\cite{Dim99} of the shell model parameters
from PDF data.\cite{DRL::Unpublished}

We calculated these peaks using a numerical implementation of Eq. (\ref{AnisotrCase:Peak})
with a step in \( r \) equal to \( 10^{-4} \) {\AA}. The maximum of each peak
(its position) is shifted to a distance larger than the corresponding average
\( R \). This shift is equal to \( 0.0186 \) {\AA} for the peak with \( R=1.956 \)
{\AA} and to \( 0.0321 \) {\AA} for the one with \( R=3.912 \) {\AA}. These
are fairly large peak shifts and, importantly, of the same magnitude as observed
peak shifts in specific materials interpreted as statitic displacements from
the periodic lattice structure. The peak shapes show deviation from the Gaussian,
as the asymmetry in the shape with respect to the peak position is clear from
the figure (a quantitative measure for this will be given below). The \( p(r) \)
decreases to zero slower for \( r>R_{max} \), where \( R_{max} \) is the distance
of the peak position, than for \( r<R_{max} \).

It is of interest to note a recent study\cite{Boz2000} on the PDF of the superconducting
La\( _{2-x} \)Sr\( _{x} \)CuO\( _{4} \) in which a change of the Cu-O bond
length equal to \( 0.024 \) {\AA} was reported. However, this bond change has
been attributed to a different mechanism. This is a layered material and it
will clearly be of importance to be able to estimate to what extent this shift
may actually be due to the effect of the powder average. 

\begin{figure}

\caption{\label{peak:shift}Contour plot of the peak shift magnitude, Eq.(\ref{def:peak:shift})
as a function of the \protect\( t_{\bot }^{-2}\protect \) and \protect\( t_{\Vert }^{-2}\protect \)
parameters.}

{\par\centering \resizebox*{2.5in}{2.5in}{\includegraphics{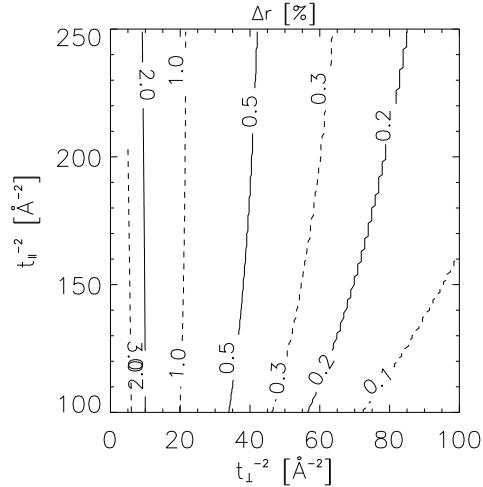}} \par}
\end{figure}

We proceed to formally study the peak shift relative to the average \( R \)
and the deviation of the peak shape from the Gaussian for the domain of \( \left( t^{-2}_{\bot },\, t^{-2}_{\Vert }\right)  \)
given above and for \( R=1.956 \) {\AA}. The peak shift\begin{equation}
\label{def:peak:shift}
\Delta r=\frac{R_{max}-R}{R}\times 100,
\end{equation}
 change relative to the average distance between the atoms in the pair is plotted
as a function of \( t_{\bot }^{-2} \) and \( t_{\Vert }^{-2} \) in Fig \ref{peak:shift}.
The figure shows that increasing the anisotropy of the eigenvalues of the \( {\mathcal{T}} \)
tensor could lead to peak shifts as large as 3\% of \( R \). Moreover, the
contour plot gives that for shifts larger than approximately 0.5\%, the value
of the \( t^{-2}_{\bot } \) remains almost unchanged on each contour level
while increasing \( t^{-2}_{\Vert } \). Note that peak shifts of the order
of 0.5\% (\( \approx 0.01 \) \AA\ for \( R \) = 1.97 \AA) or larger should
not be difficult to detect experimentally. In the limit \( t^{-2}_{\bot }\rightarrow t^{-2}_{\Vert } \)
from below, the peak shift tends to zero as it should in the isotropic limit
and the peak position is at the average distance \( R. \) 
\vspace{0.3cm}

\begin{figure}

\caption{\label{alpha::n}Contour plots of the dependence of \protect\( \alpha _{n},\, (n=2,\, 3,\, 4)\protect \)
(see Eq.(\ref{dev::Gauss})) on \protect\( \left( t^{-2}_{\Vert },\, t^{-2}_{\bot }\right) \protect \)
for the same peak as in Fig. \ref{peak:shift}.}

{\par\centering \resizebox*{2.5in}{2.5in}{\includegraphics{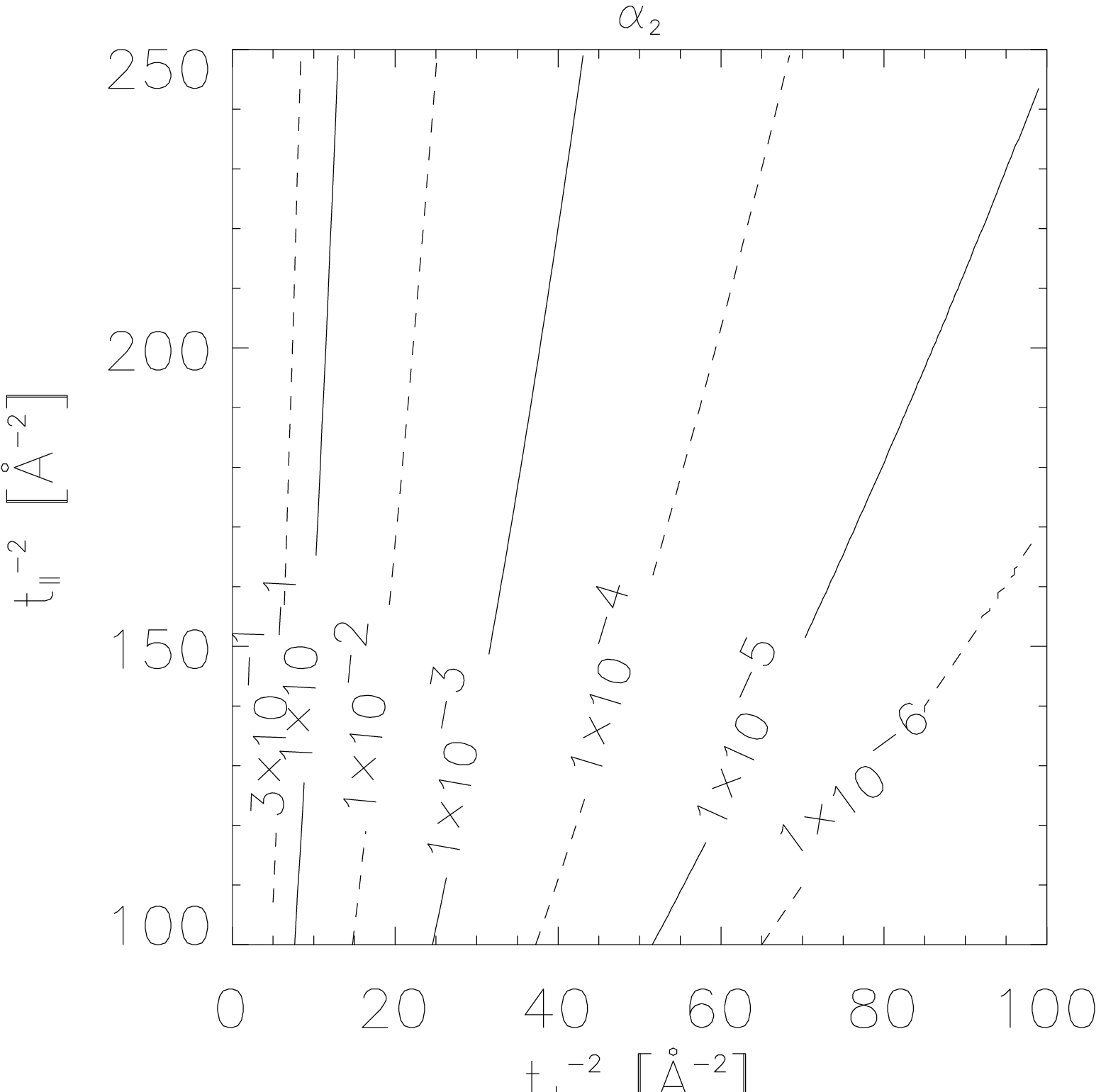}} \par}

{\par\centering \resizebox*{2.5in}{2.5in}{\includegraphics{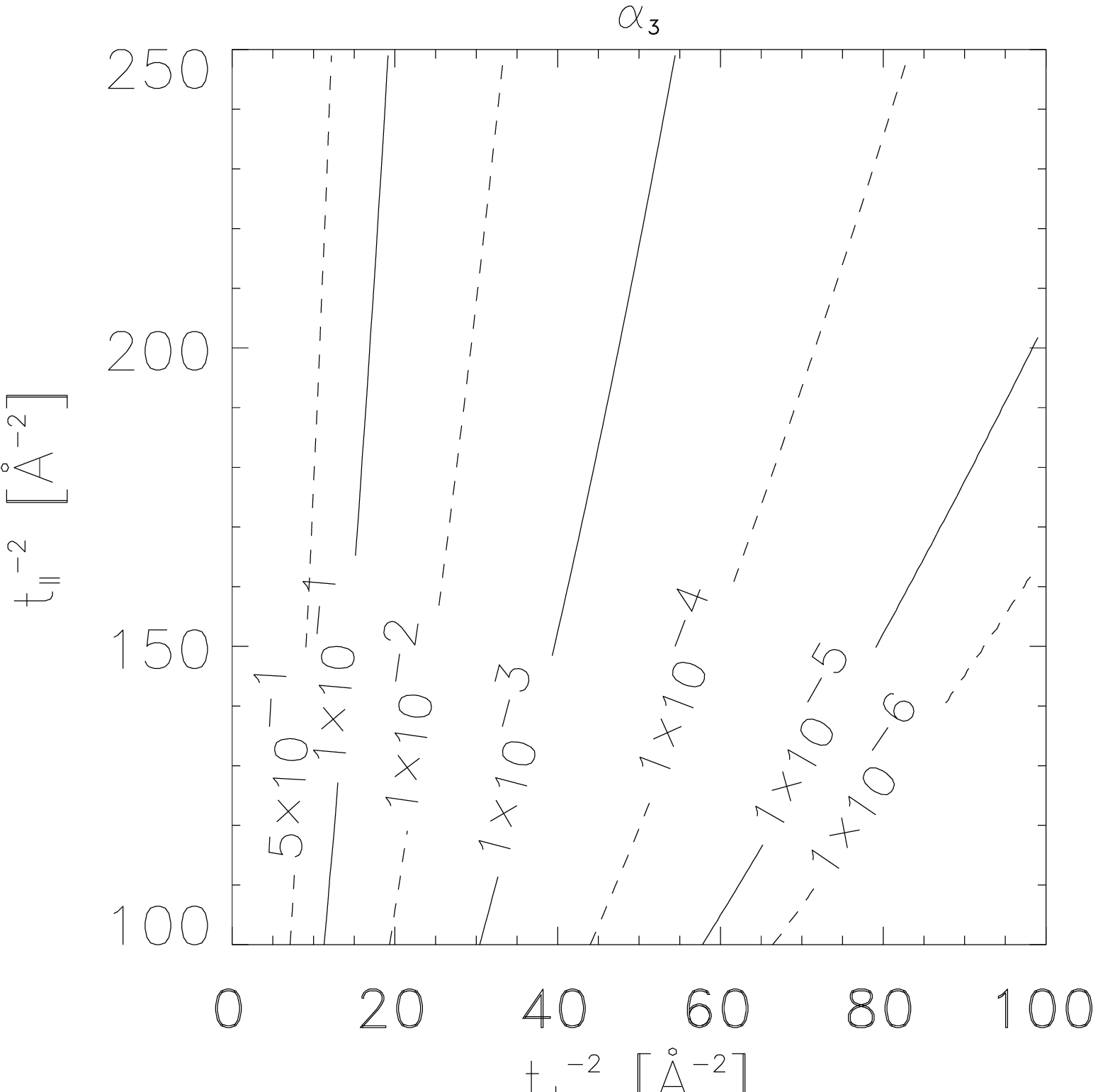}} \par}

{\par\centering \resizebox*{2.5in}{2.5in}{\includegraphics{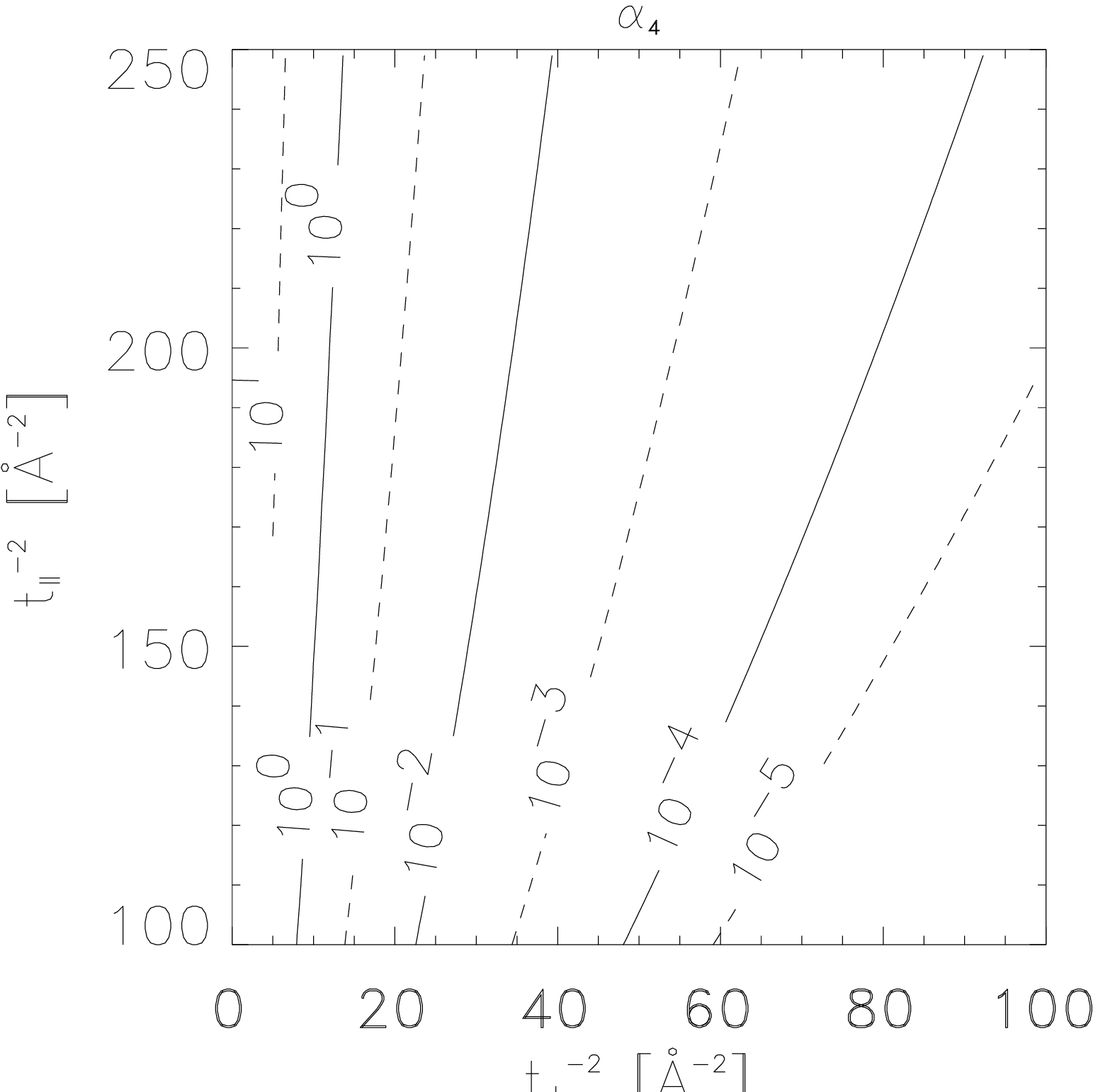}} \par}
\end{figure}

The deviation of a peak in \( r\rho (r) \) from a Gaussian form can be studied\cite{Rah64}
by calculating the moments of the normalized (to unity) distribution given by
this peak. If the normalized distribution were a Gaussian \( p_{N(\overline{r},\sigma ^{2})} \),
the moments obey\begin{eqnarray*}
\overline{\left( r-\overline{r}\right) ^{2n-1}} & = & 0,\\
\overline{\left( r-\overline{r}\right) ^{2n}} & = & (2n-1)!!\left( \overline{\left( r-\overline{r}\right) ^{2}}\right) ^{n},
\end{eqnarray*}
 for \( n=1,\, 2,\, \ldots  \).When the normalized distribution deviates from
a Gaussian, the quantity\begin{equation}
\label{dev::Gauss}
\alpha _{n}=\frac{\overline{\left( r-\overline{r}\right) ^{2n}}}{(2n-1)!!\left( \overline{\left( r-\overline{r}\right) ^{2}}\right) ^{n}}-1,\, \, \, n=2,\, 3,\, \ldots ,
\end{equation}
 will deviate from zero. The contour plots of the dependence of \( \alpha _{n} \)
(\( n=2,\, 3,\, 4 \)) on \( \left( t^{-2}_{\Vert },\, t^{-2}_{\bot }\right)  \)
for the same intervals as in Fig. \ref{peak:shift} and for the same peak is
given in Fig. \ref{alpha::n}.
The behavior of \( \alpha _{n} \) \( (n=2,\, 3,\, 4) \) as a function of \( \left( t^{-2}_{\Vert },\, t^{-2}_{\bot }\right)  \)
is very similar to the behavior of the peak shift shown in Fig. \ref{peak:shift},
as expected. When \( {\mathcal{T}} \) is close to isotropic the peak shift is
less than 0.1\%, Figs. \ref{peak:shift} and \ref{alpha::n}, which is \( \leq 0.002 \)
\AA\ for this peak, and the peak shape is close to a Gaussian \( \left( \alpha _{n}<10^{-5},\, n=2,\, 3,\, 4\right)  \).
The values of \( \alpha _{2}, \) \( \alpha _{3}, \) and \( \alpha _{4} \)
are all positive for these intervals of \( t^{-2}_{\Vert } \) and \( t^{-2}_{\bot } \)
indicating that the normalized distribution of \( r\rho (r) \) for this peak
tends to zero slower than a Gaussian with increasing \( r. \) 

The magnitude of the peak shift and the behavior of the \( \alpha _{n} \) quantities
plotted in Fig. \ref{peak:shift} and Fig. \ref{alpha::n} depend on the anisotropy
of the tensor \( {\mathcal{T}} \), e.g. determined by the ratio of its largest
to smallest eigenvalues, and on the average pair distance \( R \). Fig. \ref{peak:shift}
shows that for \( t^{-2}_{\bot }<40 \) {\AA}\( ^{-2} \), the peak position
shift is approximately constant for a constant \( t^{-2}_{\bot } \) and varying
\( t^{-2}_{\Vert } \). Thus, the peak shift of 0.82 \% for the peak with \( R=1.956 \)
{\AA} plotted in Fig. \ref{peak::shapes} and a ratio of \( t^{-2}_{\Vert }/t^{-2}_{\bot }\approx 6.17 \)
could be obtained for the same \( R \) and \( t^{-2}_{\Vert }/t^{-2}_{\bot }\approx 4 \).
Such ratios, and even much larger, have been experimentally observed at least
for the self-correlation tensors \( \left\langle \mathbf{uu}\right\rangle  \),
as can be seen from published Debye-Waller factors in highly anisotropic materials.
For example, \( \left\langle u^{2}_{\Vert }\right\rangle /\left\langle u^{2}_{\bot }\right\rangle \approx 3 \)
has been found for layered graphite crystals\cite{Layered::Graphite} and \( \left\langle u^{2}_{x}\right\rangle /\left\langle u^{2}_{z}\right\rangle \approx 19 \)
for the moniclinic 7M structure\cite{Nod90} in Ni\( _{62.5} \)Al\( _{37.5} \). 

The peak shift for a peak given by Eq. (\ref{AnisotrCase:Peak}) decreases when
increasing \( R \). This shift was \( 0.0186 \) {\AA} for the peak due to
the \( O_{\mathrm{I}} \)-\( B \) nn pair of atoms, \( \left( R,\, t^{-2}_{\bot },\, t^{-2}_{\Vert }\right) =\left( 1.956,\, 21.71,\, 133.99\right)  \),
and it reaches \( 0.001 \) {\AA} for \( R\approx 37 \) {\AA} and the same
values of \( \left( t^{-2}_{\bot },\, t^{-2}_{\Vert }\right)  \). While the
eigenvalues of the \( {\mathcal{T}} \) tensor will also change with \( R \)
due to the pair displacement correlation contributions to \( {\mathcal{T}} \),
these contributions will decrease with increasing \( R \) and for large distances,
the eigenvalues of \( {\mathcal{T}} \) will be approximately given by the displacement-displacement
self correlations. Then for very anisotropic Debye-Waller factors \( {\mathcal{T}} \)
will also be very anisotropic and peak shifts will result for the powder average.
More importantly, these shifts will be different for the atomic pairs at different
distances with shifts as large as e.g. \( 0.02 \) {\AA} for the nn bonds and
decreasing slowly to \( 0.001 \) {\AA} for \( R\approx 37 \) {\AA}. Moreover,
these shifts are confined to the pairs in which at least one of the atoms is
O, in our perovskite example, while peaks due to A-A or B-B pairs could be positioned
very close to their respective average \( R \)'s at all distances, because
for these atoms the Debye-Waller factors are very close to isotropic. Thus,
we may have a crystal with several different types of atoms in the unit cell,
very well defined periodic structure, and nevertheless some of the peak positions
in its powder averaged PDF could deviate one or more percent from the average
atomic pair distances while other peaks could be practically positioned, within
the resolution in \( r \) space, at the average \( R \)'s.

\section{Summary}

\label{Implications}

We have considered the effect of the powder average on the peak shapes and positions
of the pair distribution function as measured from neutron diffraction, assuming
that the harmonic approximation is a valid description for the phonons in the
system. When the pair correlation displacement tensor \( {\mathcal{T}}(ld;l'd') \)
for a given pair of atoms is isotropic the shape of the peak in \( r\rho (r) \)
(and \emph{not} in \( \rho (r) \)) corresponding to this pair is a Gaussian
scaled by the average distance \( R(ld;l'd') \) between the atoms and a factor
which depends on the scattering lengths. However, for polycrystalline samples,
the pair correlation tensor \( {\mathcal{T}}(ld;l'd') \) is generally not isotropic.
Then the powder average leads to deviations from the Gaussian shape and to peak
shifts depending on the {}``degree of anisotropy{}'' of \( {\mathcal{T}} \)
and the magnitude of the average distance \( R. \) The anisotropy of \( {\mathcal{T}} \)
depends on the relative magnitude of its eigenvalues. For the specific \( {\mathcal{T}} \)
and \( {{\mathbf{R}}} \) of a pair of atoms, the magnitude of the peak shift can
be as large as several percent of the of the magnitude of the average pair distance
\( R \). This could be of the order of 0.02 \AA\ or larger. Such a change in
a peak position should be possible to detect experimentally given the currently
available, high resolution neutron diffraction measurements using high intensity
neutron pulse sources. Analysis of the moments of the normalized distribution
of a peak in \( r\rho (r) \) on the anisotropy of \( {\mathcal{T}} \) shows
that the normalized distribution tends to zero slower than a Gaussian for increasing
r. It is also possible to have a highly anisotropic material for which some
of the peaks in its powder averaged PDF are displaced markedly from the average
positions while other peaks are at their atomic pairs average distances. In
this case, the Rietveld analysis of the Bragg peaks may show that the crystal
is periodic. 

Generally, for any experiment which allows to obtain the powder averaged
$\rho(r)$ accurately enough for (perfect) harmonic crystals in order to
make a meaningful comparison with the theoretical PDF as obtained from 
Eqs. (\ref{PDF:HA}) and (\ref{PDF:Powder}), 
the peak shifts and shapes reported here should
be relevant effects. We considered in this paper the case of neutron
powder diffraction experiments which we expect to be an appropriate 
probe to detect the peak shape and position effects we predict based on
the theoretical results. Other experiments which allow to probe the PDF of
a polycrystalline material are x-ray scattering and EXAFS.\cite{Kon88} 
Similarly to the neutron experiments, a structure factor can be deduced from
x-ray scattering data. In principle, from its inversion 
(Eq. (\ref{Rhor::via::Sq})), a
PDF can be obtained. This is, however, more complicated for x-rays than for
neutrons since the scattering lengths for x-rays (the atomic form factors) are 
q-dependent rather than just numbers in the case of neutrons. 
Powder averaged PDF are often used in the interpretation
of EXAFS spectra but Gaussian peaks are assumed at the average distances. 

Clearly, experimental confirmation of our result will be very important for
the proper analysis and understanding of PDF data in very 
anisotropic materials.

Work at Los Alamos is supported by
the Department of Energy under contract W-7405-ENG-36.

\section*{Acknowledgements}

D. A. Dimitrov would like to express his gratitude to 
Professors T. Egami and D. Louca 
for valuable discussions regarding the neutron diffractoin PDF experiments.


\begin{thebibliography}{10}
\bibitem{Lif52}I. M. Lifshitz, Zh. Eksp. Teor. Fiz. \textbf{22}, 475 (1952). 
\bibitem{Kri82}M. A. Krivoglaz and A. V. Min'kov, Sov. Phys. Crystallogr. \textbf{27}, 503
(1982) and the references therein. 
\bibitem{Hov54}L. Van Hove, Phys. Rev. \textbf{95}, 249 (1954).
\bibitem{Was80}Y. Waseda, \emph{The Structure of Non-Crystalline Materials} (McGraw-Hill, New
York, 1980).
\bibitem{Rah62::1}A. Rahman, K. S. Singwi, and A. Sj\"olander, Phys. Rev. \textbf{126}, 986 (1962).
This paper approximates the self correlation function \( G_{s}\left( r,\, t\right)  \)
with a Gaussian which is justified for very short and long times. 
\bibitem{Kap65}R. Kaplow, S. L. Strong, and B. L. Averbach, Phys. Rev. \textbf{138}, A 1336
(1965). The radial distribution function obtained from X-ray diffraction scattering
data is modeled with gaussians. 
\bibitem{Ega90}T. Egami, Mater. Trans. {\bf 31}, 163 (1990).
\bibitem{Tob91}B. H. Toby and T. Egami, Acta Cryst. A {\bf 48}, 336 (1991).
\bibitem{Squ96}See, e.g., G. L. Squires, \emph{Introduction to the Theory of Thermal Neutron
Scattering} (Cambridge University Press, New York, 1978).
\bibitem{Pla52}G. Placzek, Phys. Rev. \textbf{86}, 377 (1952).
\bibitem{Wic54}G. C. Wick, Phys. Rev. \textbf{94}, 1228 (1954).
\bibitem{Yar73}J. L. Yarnell, M. J. Katz, R. G. Wenzel, and S. H. Koenig, Phys. Rev. A \textbf{7},
2130 (1973).
\bibitem{Asc66}P. Ascarelli and G. Caglioti, Nuovo Cimento \textbf{43B}, 376 (1966).
\bibitem{Ege87}P. A. Egelstaff, in \emph{Methods of Experimental Physics} v. 23\emph{,} Part
B, \emph{Neutron Scattering}, edited by D. L. Price and K. Sk\"old (Academic
Press, San Diego, 1987).
\bibitem{Dim99}D. A. Dimitrov, D. Louca, and R\"oder, Phys. Rev. B {\bf 60}, 6204 (1999).
\bibitem{Lov85v1}S. W. Lovesey, {\it Theory of Neutron Scattering from Condensed Matter} (Clarendon
Press, Oxford, 1985), v. 1.
\bibitem{BH54}M. Born and K. Huang, {\it Dynamical Theory of Crystal Lattices}, (Oxford University
Press, Oxford, England, 1954).
\bibitem{Cow64}R. A. Cowley, Phys. Rev. B {\bf 134}, A981 (1964).
\bibitem{Mig76}R. Migoni, H. Bilz, and D. B\"auerle, Phys. Rev. Lett. {\bf 37}, 1155 (1976).
\bibitem{DRL::Unpublished}D. A. Dimitrov, H. R\"oder, D. Louca, unpublished.
\bibitem{Boz2000}E. S. Bozin, G. H. Kwei, H. Takagi, and S. J. L. Billinge, Phys. Rev. Lett.
\textbf{84}, 5856 (2000).
\bibitem{Rah64}A. Rahman, Phys. Rev. A \textbf{136}, 405 (1964).
\bibitem{Layered::Graphite}G. Albinet, J. P. Biberian, and M. Bienfait, Phys. Rev. B \textbf{71}, 2015
(1971); Von A. Ludsteck, Acta Cryst. \textbf{A28}, 59 (1972); J. P. Biberian
and M. Bienfait, Acta Cryst. \textbf{A29}, 221 (1973).
\bibitem{Nod90}Y. Noda, S. M. Shapiro, G. Shirane, Y. Yamada, and L. E. Tanner, Phys. Rev.
B \textbf{42}, 10397 (1990).
\bibitem{Kon88}\emph{X-Ray Absorption, Principles, Applications, Techniques of EXAFS, SEXAFS
and XANES}, edited by D. C. Konigsberger and R. Prins (Wiley, New York, 1988).\end{thebibliography}
\end{document}